\documentclass[12pt, twocolumn]{IEEEtran}
\usepackage{graphicx}
\usepackage{epstopdf}
\usepackage{amsmath}
\usepackage{amssymb}
\usepackage{cite}
\usepackage{color}
\usepackage{amsthm}
\theoremstyle{plain}	 % definition

\usepackage{mathtools}
\usepackage[normalem]{ulem}
\usepackage{cancel}
\usepackage{cite}
\usepackage{textcomp}
\usepackage{physics}
\usepackage{amsfonts}
\usepackage{bbold}
\usepackage{enumitem}
\usepackage{threeparttable}
\usepackage{multirow}
\usepackage{booktabs}

\newcommand\blfootnote[1]{%
  \begingroup
  \renewcommand\thefootnote{}\footnote{#1}%
  \addtocounter{footnote}{-1}%
  \endgroup
}

\newcommand{\mc}[1]{\mathcal{#1}}

\newcommand\numberthis{\addtocounter{equation}{1}\tag{\theequation}}
\newcommand{\mb}[1]{\mathbb{#1}}

\begin{document}
\title{Multi-Hop Routing in Covert Wireless Networks}

\author{\IEEEauthorblockN{Azadeh Sheikholeslami\IEEEauthorrefmark{1},
		Majid Ghaderi\IEEEauthorrefmark{2},
		Don Towsley\IEEEauthorrefmark{3},
		Boulat A. Bash\IEEEauthorrefmark{4},\\
		Saikat Guha\IEEEauthorrefmark{5}},
	Dennis Goeckel\IEEEauthorrefmark{1}\\
	\IEEEauthorblockA{\IEEEauthorrefmark{1}Dept. of Elec. and Comp. Engineering (ECE), Univ. of Massachusetts, Amherst%University of Massachusetts, Amherst%, MA %01003--9292
	}\\
	\IEEEauthorblockA{\IEEEauthorrefmark{2} Department of Computer Science , University of Calgary%, Calgary %01003--9292
	}\\
	\IEEEauthorblockA{\IEEEauthorrefmark{3}College of Info. and Comp. Sciences (CICS), Univ. of Massachusetts, Amherst%, MA %01003--9264
	}\\
	\IEEEauthorblockA{\IEEEauthorrefmark{4} Raytheon BBN Technologies, Cambridge, MA %02138
		}\\
	\IEEEauthorblockA{\IEEEauthorrefmark{5} The Univ. of Arizona, College of Optical Sciences, Tucson, AZ%, Tucson, AR %01003--9264
	}}

\maketitle
\thispagestyle{empty}
\begin{abstract}
	In covert communication, Alice tries to communicate with Bob without being detected by a warden Willie. 
When the distance between Alice and Bob becomes large compared to the distance between Alice and Willie(s), the performance of covert communication will be degraded.
In this case,  multi-hop message transmission via intermediate relays can help to improve  performance.
Hence, in this work multi-hop covert communication over a moderate size network and in the presence of multiple collaborating Willies is considered.
The relays can transmit covertly  using either  a single key for all relays, or  different independent keys at the relays. 
For each case, we develop efficient algorithms to find optimal paths with maximum throughput and minimum end-to-end delay between Alice and Bob.
As expected, employing multiple hops significantly improves the ability to communicate covertly versus the case of a single-hop transmission.  Furthermore, at the expense of more shared key bits,  analytical results
 and  numerical simulations demonstrate that multi-hop covert communication with different independent keys at the relays has better performance than multi-hop covert communication with a single key.

\end{abstract}

\blfootnote{This work was sponsored by the National Science Foundation (NSF) under grants ECCS-1309573 and CNS-1564067, and DARPA under contract number HR0011-16-C-0111.}

\section{Introduction}\label{sec:introduction}

Due to the broadcast nature of wireless networks, any node near a transmitter can overhear the message.
Thus, providing  security for  wireless communications is of central importance   and has attracted particular attention. 
Various security schemes have been developed to protect the content of a message from an unintended recipient\cite{paar2010understanding,wyner1975wire,cachin1997unconditional,sheikholeslami2013everlasting,sheikholeslami2017energy,sheikholeslami2015jamming};
however, there are security scenarios where the \textit{existence} of a transmission (or the transmitter) is  to be kept hidden from adversaries.
In such adversarial scenarios,  traditional security approaches are no longer effective, and the communicating parties should seek low probability of detection approaches, which have been studied recently and termed ``covert communication"  \cite{bash2013limits,che2013reliable,soltani2014covert,bloch2015covert,bash2015quantum,wang2016fundamental,bash2016covert,sheikholeslami2016covert,goeckel2016covert,he2017covert,hu2017covert}.
Consider a wireless communication scenario when Alice (the transmitter) wants to send a message to Bob (the intended receiver) such that an attentive adversary Willie is not aware of the transmission.
In \cite{bash2013limits}, it is shown that  using a pre-shared key between Alice and Bob, it is possible to transmit $\mc{O}(\sqrt{n})$ information bits covertly over $n$ channel uses  such that Willie is not aware of the existence of communication. Moreover, it is not possible to transmit $\omega(\sqrt{n})$ bits over $n$ channel uses covertly: if the transmitter transmits $\omega(\sqrt{n})$ bits either Willie can detect the communication, or Bob will not be able to decode the message with a (arbitrarily) low probability of error.
%Covert communication over an slotted channel \cite{}, when Willie does not know the background noise power \cite{}, and over quantum channels  

In  \cite{bloch2015covert,wang2016fundamental} the constant in front of $\sqrt{n}$ for the number of bits transmitted covertly over memoryless channels is characterized. It is shown that the number of bits that can go through the channel without being detected by Willie has a direct relationship with the distance between the probability distribution functions of the received signals at Bob when no communication occurs and when Alice is transmitting. Also, it has an inverse relationship with the distance between the probability distribution functions of the received signals at Willie when no communication occurs and when Alice is transmitting \cite{bloch2015covert,wang2016fundamental}.

In an environment with  AWGN channels when Alice and Bob are located far from each other, in order to make the probability of error at Bob sufficiently small, Alice should use a high  transmit power. 
However, this  increases the probability of being detected by Willie, especially if Willie is close to Alice and thus receives a strong signal, and/or if multiple collaborating Willies are present and try to detect any transmission.
In order to solve this problem, in \cite{soltani2014covert} Alice and Bob  use artificial noise from friendly chatterers to  increase the noise level of the wireless environment to help them hide their communication even when Willie is close to Alice and when multiple collaborating Willies are present.
However, this approach requires  some friendly system nodes in the network who are not concerned to transmit openly and reveal their existence and/or their locations. Hence,  when  all system nodes   prefer to hide their existence and/or their locations, the scheme of \cite{soltani2014covert} cannot be used.
In this case,  in order to facilitate covert communication, we propose utilizing the friendly system nodes to establish a multi-hop path from Alice to Bob. On this path, the distance between  intermediate relays  is short and thus each relay can transmit with a small transmit power in order to decrease  its probability of being detected. Also, the multi-hop path from Alice to Bob can take detours to avoid Willies. That is, the routing algorithm can be designed so that it chooses relays that are less susceptible to being detected by Willies.

In this paper, we consider multi-hop covert communication between Alice and Bob  in the presence of multiple Willies. 
In order to consider the most powerful adversary scenario, we assume all Willies are collaborating to detect any transmission of Alice  and the intermediate relays.
A message  generated by Alice  travels hop-by-hop until it is delivered to Bob.
For covert communication, Alice  and the intermediate relays  use a key to encode the message. We consider two scenarios. In the first scenario, we consider the case that a single key is used by Alice and all relays at all hops to encode the message.
%We can either use a single key to encode the message at all relays, or use independent keys at the relays.
%At each hop the receiving  relay  de-noises the received signal and forwards it to the next relay. 
While this approach is simple and does not require separate keys at each hop, it can increase the probability of being detected  because the exact same codeword is transmitted over multiple links and is observed by Willies. 
%This will degrade the performance of the proposed algorithms. 
In the second scenario, we  consider employing  independent keys at the relays. Each relay encodes the received message with an independent  key and then forwards it to the next relay. In this case, independent codewords are transmitted over different links, and thus the signals being observed by the Willies at different hops are independent.

We consider two performance metrics, namely, throughput and end-to-end delay over the path from Alice to Bob. We develop algorithms to find the path with the maximum throughput and the path with minimum  end-to-end delay between Alice and Bob for  the case of a single key and the case of independent keys at the relays. We compare the performances of all algorithms numerically as we vary the network parameters.

The rest of this paper is organized as follows. In Section \ref{sec:system} a short summary of covert communications,  the system model, the covertness criteria, and the multi-hop strategies used in this work are explained. In Sections \ref{sec:AF} and \ref{sec:DF}, multi-hop covert communication with a single key and with independent keys at the relays are considered, respectively, and for each case algorithms to establish an optimal path from Alice to Bob are proposed. The proposed algorithms are studied and compared numerically in Section \ref{sec:numerical}. Concluding remarks are discussed in Section \ref{sec:conc}.

\section{Prerequisites}\label{sec:system}

\subsection{Covert Communication or Communication with Low Probability of Detection}
Consider a transmitter Alice, a receiver Bob, and a warden Willie.  Alice wants to transmit a message to Bob such that Willie is not aware of the communication. Willie uses his observations of the channel to detect whether Alice transmits or not. Suppose $H_1$ is the hypothesis that Alice transmits a signal, and $H_0$ is the hypothesis that no communication occurs. Willie's probability of detection error   consists of two components: the probability of missed detection (Willie declares no communication  when Alice transmits) denoted by $\mathbb{P}^W_{MD}=\mathbb{P}(H_0|H_1 \;\text{is correct})$, and the probability of false alarm (Willie declares  communication  when no communication takes place) denoted by $\mathbb{P}^W_{FA}=\mathbb{P}(H_1|H_0\; \text{is correct})$. Hence, considering equal prior probabilities, the total detection error of Willie is:
\begin{eqnarray}
\mathbb{P}^W_e=\frac{\mathbb{P}^W_{FA}+\mathbb{P}^W_{MD}}{2}.
\end{eqnarray}

In covert communication,  the goal is to prevent Willie from using his observations of the channel to make the probability of detection error $\mathbb{P}^W_e$ arbitrarily small.
In order to reach this goal, Alice and Bob pre-share a secret key, based on which Alice selects a codebook from an ensemble of codebooks.  Assume that the  channel between Alice and Willie experiences some sort of uncertainty (e.g. it is an AWGN channel). The codebooks that Alice chooses from are  low power codebooks such that  Willie, without knowing the key, cannot decide with arbitrarily low probability of detection error that whether  his observation is a signal transmitted by Alice or a result of the uncertainty of the channel. 
In  \cite{bash2013limits}, it is shown that  the power of the signal transmitted over $n$ channel uses should be of order of $\frac{1}{\sqrt{n}}$, which allows transmission of $\mc{O}(\sqrt{n})$ covert bits  over $n$ channel uses. 
Note that unlike conventional communication, in covert communication throughput changes with the number of channel uses $n$, and is on the order of $\frac{1}{\sqrt{n}}$.
In \cite{bloch2015covert}, it is shown that for covert communication the number of key bits shared between Alice and Bob should be on the order of $\sqrt{n}$, and using this key, Bob can decode the message with arbitrarily low probability of error.

\subsection{System Model}\label{sec:model}

We consider a wireless network that consists of multiple (friendly) system nodes which are  distributed arbitrarily.
The set of (friendly) system nodes is denoted by $\mc{T}=\{T_1,\ldots,T_N\}$, where $N$ is the number of such nodes in the network.
In addition to the system nodes, multiple collaborating Willies, i.e. the wardens that want to detect any communication in the network, are  present. The set of Willies is denoted by $\mc{W}=\{W_1,\ldots,W_M\}$, where $M$ is the number of Willies.
Willies collaborate and use all of their observations (across Willies, across transmissions and across time) to  attempt to determine whether  any of the system nodes transmitted or not. 

%We assume that the locations of  Willies are known to the system nodes, and the locations of the  system nodes are known to Willies.
 The channel between nodes is an additive white Gaussian noise (AWGN) channel with path-loss exponent $\alpha$, where $\alpha=2$ corresponds to free space, and $\alpha>2$ corresponds to a terrestrial environment. 
 Any  transmitter $X$ in the network attempts to transmit a message by employing a Gaussian codebook \cite{bash2013limits}, and its transmitted signal is given by $[f_1, f_2, \ldots, f_n]$, where $f_j\sim\mc{N}(0,1)$ and $n$ is the length of each codeword.
 %Suppose a relay $S(\ell_i)$ along the path between Alice and Bob transmits the message $[f_1, f_2, \ldots, f_n]$ to  relay $D(\ell_i)$. 
 The signal that  a receiver $Y$ receives is,
 \begin{equation}
 Z_j^{({Y})}=\frac{\sqrt{P_X}f_j}{d_{X,Y}^{\alpha/2}}+N_j^{({Y})},\; j=1,\ldots,n,
 \end{equation}
 where $P_X$ is the transmit power of node $X$, $d_{X,Y}$ is the distance between transmitter $X$ and receiver $Y$, and $N_j^{({Y})}\sim\mc{N}(0,\sigma_{Y}^2)$ is AWGN at the receiver.
The signal that Willie $W_k$ receives is,
  \begin{equation}
  Z_j^{(W_k)}=\frac{\sqrt{P_X}f_j}{d_{X,W_k}^{\alpha/2}}+N_j^{(W_k)},\; j=1,\ldots,n,
  \end{equation}
where $d_{X,W_k}$ is the distance between the transmitter $X$ and Willie $W_k$, and $N_j^{(W_k)}\sim\mc{N}(0,\sigma_{W_k}^2)$ is AWGN at Willie $W_k$.
Throughout this paper, it is assumed that the distances between the system nodes and the distances between the system nodes and the Willies are known to the system nodes and the Willies. In the case that the knowledge of the locations of the Willies is not complete, we can use lower bounds on the distances between the transmitters and the Willies to obtain bounds on the allowable transmit  powers such that  Willies  cannot detect  the communication (similar to the analysis presented in \cite{goeckel2016covert}).

\subsection{Covertness Criteria and Covert Throughput}\label{sec:metric}
 A transmission in the presence of  Willies is considered covert when for any $\epsilon>0$, the sum of probabilities of detection errors of their joint decision is lower bounded as,
	\begin{equation}\label{eq:covert1}
	\mathbb{P}^W_{FA}+\mathbb{P}^W_{MD}\geq 1-\epsilon,
	\end{equation}
	for sufficiently large $n$  (recall that $n$ is the length of the codewords) \cite{bash2013limits}. 
	The joint probability distribution function of  Willies' observations when a transmission occurs is given by $\mb{Q}_1$, and the joint probability distribution function of  Willies' observations when no transmission occurs is given by $\mb{Q}_0$.
	Suppose Willies perform the optimal test.
	Thus, using Pinsker's inequality \cite{lehmann2006testing,cover2012elements},
		\begin{equation}\label{eq:covert2}
		\mathbb{P}^W_{FA}+\mathbb{P}^W_{MD}\geq 1-\sqrt{\frac{1}{2}\mb{D}(\mathbb{Q}_1\|\mathbb{Q}_0)},
		\end{equation}
where $\mb{D}(\mathbb{Q}_1\|\mathbb{Q}_0)$ is the relative entropy between $\mathbb{Q}_1$ and $\mathbb{Q}_0$. Hence, combining (\ref{eq:covert1}) and (\ref{eq:covert2}), an alternative covertness criteria is to bound the relative entropy:
\begin{equation}
\mb{D}(\mathbb{Q}_1\|\mathbb{Q}_0)\leq\delta,\label{eq:covert3}
\end{equation}
where $\delta=2\epsilon^2$.
That is, if $\mathbb{Q}_1$ and $\mathbb{Q}_0$ are such that  $\mathbb{D}(\mathbb{Q}_1\|\mathbb{Q}_0)\leq\delta$, it is guaranteed that the communication is covert, i.e. $\mathbb{P}^W_{FA}+\mathbb{P}^W_{MD}\geq 1-\epsilon$. In this paper, we consider (\ref{eq:covert3}) as our covertness criteria.

A transmission from a transmitter $X$ to a receiver $Y$ is considered reliable if as the block-length $n$ goes to infinity, the average error probability of  receiving a message at  receiver $Y$  approaches  zero.
We define  \textit{covert throughput}  as the rate of reliable communication between a transmitter $X$ and a receiver $Y$ such that the communication is hidden from warden Willies. 
In this paper, we use the terms ``throughput'' and ``covert throughput'' interchangeably. 
For an AWGN channel, it has been shown that if the transmitter uses zero mean Gaussian  input symbols with average power $P_X$, any covert throughput less than 
\begin{eqnarray}
C=\frac{1}{2}\log\left( 1+\frac{P_X}{\sigma_{Y}^2d_{X,Y}^{\alpha}}\right) ,
\end{eqnarray}
can be achieved reliably \cite{cover2012elements}. Note that $C$  depends on the Willies' distances to the transmitter through $P_X$.
As mentioned before, 
in order to guarantee covertness we should have $P_X=\mc{O}(\frac{1}{\sqrt{n}})$. 
Since $P_X$ becomes very small  as $n$ tends to $\infty$, the approximation
\begin{eqnarray*}
	C\approx\frac{P_X}{2\sigma_{Y}^2d_{X,Y}^{\alpha}},
\end{eqnarray*}
is tight for large $n$, and hence we will employ
\begin{eqnarray}\label{eq:rate}
	C=\frac{P_X}{2\sigma_{Y}^2d_{X,Y}^{\alpha}},
\end{eqnarray}
for our network design.

\subsection{Multi-Hop  Strategies}
As mentioned in Section \ref{sec:introduction}, in order to improve the performance of communication between Alice and Bob,  we consider multi-hop transmission. 
Alice, the source node,  transmits a message to Bob, the destination node, in a multi-hop fashion. 
Let a path from Alice to Bob be denoted by $\Pi=\left( \ell_1,\ldots,\ell_H\right) $, where $H$ is the number of hops of the path from Alice to Bob, and $\ell_i=\left( S(\ell_i),D(\ell_i)\right)$ is the link between node $S(\ell_i)$ and node $D(\ell_i)$ along the path, where $S(\ell_1)$ is Alice and $D(\ell_H)$ is Bob. 

We consider two approaches: multi-hop communication \textit{with a single key (SK)} and multi-hop communication \textit{with independent keys at the relays (IK)}. 
With a single key, all  relays use the same key to encode the message, i.e. a message received from the previous relay is  sent to the next relay using the same key. 
Hence, the exact same codeword is transmitted over every hop on the path from Alice to Bob. On the other hand, with independent keys at the relays, the message is re-encoded at each hop with a different key sequence such that the codeword sent over each link is independent of the codewords sent over other links of the path.
For each  approach, we first optimize transmission along  a given path $\Pi$ between Alice and Bob. 
In particular, we find the optimal powers that should be allocated to each relay along the path such that the end-to-end covertness constraint is satisfied, and the desired performance metric (covert throughput or end-to-end delay) is optimized. 
Then,   for each case, we exploit these results to develop a routing algorithm that not only allocates the optimal powers to the relays, but also finds the optimal  path  $\Pi^*$ from the set $\mathbf{\Pi}$ of all possible paths between Alice and Bob. 

\subsection{Key Distribution}\label{sec:key}

Considering the fact that we need $\mc{O}(\sqrt{n})$ number of key-bits to encode each message, it seems quite challenging to exchange such a long key sequence in an adversarial environment. 
In particular, for multi-hop covert communication with independent keys at the relays (IK) many  such long key sequences are needed.
Fortunately, in practice (very) short  key sequences shared between the relays are sufficient to generate the (very) long key sequences required for covert communication.
As described  in  \cite[Section II]{sheikholeslami2013everlasting},   the relays can use the short key sequences as the initial keys  for a stream-cipher generating scheme to generate the  long key sequences. For instance, a stream-cipher generating scheme called   Trivium \cite{robshaw2008new} with an 80-bit initial key  can generate a $2^{64}$-bit key sequence  \cite{sheikholeslami2013everlasting}. 

\section{Covert Communication with a Single Key (SK)}
\label{sec:AF}
In this section, we consider multi-hop covert communication with a single key. Consider an $H$-hop path $\Pi=\left( \ell_1,\ldots,\ell_{H}\right)$ between Alice and Bob. Every  relay $S(\ell_i)$  forwards the message  to the next relay $D(\ell_i)$ using the same key until it is delivered to the destination, Bob. 

Consider an arbitrary Willie $W_k$ observing the  message transmission over $\Pi$. Since $W_k$ can observe the transmission of every relay along the path, it can use its observations across hops to decide whether a transmission occurs or not. This is equivalent to the case that $H$ cooperating  Willies $W_{k_1},\ldots,W_{k_H}$ are present at the location of Willie $W_k$, such that $W_{k_i}$ monitors the transmission of only the $i^{th}$ hop. Then, $W_{k_1},\ldots,W_{k_H}$ use their observations to constitute the total observation of Willie $W_k$ over all hops. Hence,   Willie $W_k$'s observations of the $i^{th}$ hop ($k=1,\ldots, M$ and $i=1,\ldots, H$) under hypothesis $H_1$ are described as:
\begin{align}
Z^{(i,k)}_{j} = \frac{\sqrt{P_i}f_{i,j}}{d_{i,k}^{\alpha/2}} + N^{(W_k)}_j,\; j=1,2, \ldots, n,
\end{align}
where $P_i$ is the transmit power of relay $S(\ell_i)$,  $f_{i,j}$ is the symbol that is sent over the $i^{th}$ hop during the $j^{th}$ symbol period, and $d_{i,k}$ is the distance from relay $S(\ell_i)$ to Willie $W_k$. Since, in this case, the same key is used to encode the message at every relay, the same symbol is sent over different hops ($f_{i,j}=f_{j}$) and thus, under hypothesis $H_1$, we have,
\begin{align}
Z^{(i,k)}_{j} = \frac{\sqrt{P_i}f_{j}}{d_{i,k}^{\alpha/2}} + N^{(W_k)}_j,\; j=1,2, \ldots, n.
\end{align}
Under hypothesis $H_0$, the Willies observations are given by,
\begin{align}
Z^{(i,k)}_{j} = N^{(W_k)}_j,\; j=1,2, \ldots, n.
\end{align}
Willies, with their collective observations over all Willies ($k$), hops ($i$) and  symbol periods ($j$), attempt to detect message transmission.

\subsection{Covertness Analysis of Covert Communication with a Single Key}\label{sec:covertness_AF}\label{sec:cov_af}

Suppose $\mathbb{Q}_0$ is the joint probability distribution   of Willies' observations over $M$ Willies, over $H$ hops, and over $n$ channel uses under hypothesis $H_0$,
and  $\mathbb{Q}_1$ is the joint probability distribution of Willies' observations over $M$ Willies,  over $H$ hops and over $n$ channel uses under hypothesis $H_1$.
Hence,  $\mathbb{Q}_0$ is a zero-mean multivariate Gaussian probability distribution function  with  covariance matrix
\begin{eqnarray}
\Sigma_0=S\otimes I_{n\times n},
\end{eqnarray}
where $S$ is an $HM\times HM$ diagonal matrix
\begin{eqnarray}
S=\text{diag }(\underbrace{\sigma_{W_1}^2,\ldots,\sigma_{W_1}^2}_{H\; \text{times}},\ldots,\sigma_{W_M}^2 ).
\end{eqnarray}
Note that each $\sigma_{W_k}^2,\:k=1,\ldots,M$ is repeated $H$ times in $S$ because each Willie $W_k$ observes the transmission of the same message over all $H$ hops.
 On the other hand, $\mathbb{Q}_1$ is  a zero-mean multivariate Gaussian probability distribution function with  covariance matrix \begin{eqnarray}
\Sigma_1=(S+UU^T)\otimes I_{n\times n},
\end{eqnarray} 
where $U$ is a column vector with $HM$ elements,
\begin{eqnarray}
U=\left[ \frac{\sqrt{P_{1}}}{d_{1,1}^{\alpha/2}},\ldots,\frac{\sqrt{P_{H}}}{d_{H,1}^{\alpha/2}},\ldots,\frac{\sqrt{P_{1}}}{d_{1,M}^{\alpha/2}},\ldots,\frac{\sqrt{P_{H}}}{d_{H,M}^{\alpha/2}}\right] ^T.
\end{eqnarray}
Suppose the Willies apply the optimal hypothesis test. 
Since $\mathbb{Q}_1$ and $\mathbb{Q}_0$ are multivariate Gaussian distributions, the relative entropy between them is given by (Appendix \ref{ap:0}),
\begin{align}\label{eq:rel_en_multi}
\nonumber\mathbb{D}&(\mathbb{Q}_1\|\mathbb{Q}_0)= \frac{1}{2}\Big( \Tr(\Sigma_0^{-1}\Sigma_1)-\dim(\Sigma_0)\\
&-\ln\frac{|\Sigma_1|}{|\Sigma_0|} +(\mu_0-\mu_1)^T\Sigma_0^{-1}(\mu_0-\mu_1)\Big),
\end{align}
where $\mu_0$ is the mean of $\mb{Q}_0$,   $\mu_1$ is the mean of $\mb{Q}_1$, $|\Sigma_0|$ is the determinant of $\Sigma_0$, and $\dim(\Sigma_0)$ is dimension of $\Sigma_0$.
Replacing $\mu_0, \mu_1, \Sigma_0$ and $\Sigma_1$ in (\ref{eq:rel_en_multi}) and performing some algebraic manipulations (Appendix \ref{ap:1}), the relative entropy in (\ref{eq:covert2}) can be written as,
\begin{align}\label{eq:D}
&\nonumber\mathbb{D}(\mathbb{Q}_1\|\mathbb{Q}_0)\\
&=\frac{n}{2}\Bigg(\sum_{\substack{\ell_i\in\Pi\\ W_k\in\mc{W}}} \frac{P_i} {\sigma^2_{W_k}d_{i,k}^{\alpha}}-\ln\Big(1+\sum_{\substack{\ell_i\in\Pi\\ W_k\in\mc{W}}}\frac{P_i}{\sigma^2_{W_k}d_{i,k}^{\alpha}}\Big)\Bigg).	
\end{align}
Using the inequality $\ln(1+x)\geq x-\frac{x^2}{2}$ for $x\geq 0$, 
\begin{eqnarray}\label{eq:DD}
\mathbb{D}(\mathbb{Q}_1\|\mathbb{Q}_0)\leq\frac{n}{4}\left(\sum_{\ell_i\in\Pi}\sum_{W_k\in\mc{W}}\frac{P_i}{\sigma^2_{W_k}d_{i,k}^{\alpha}}\right)^2.
\end{eqnarray}
Combining (\ref{eq:covert3}) and (\ref{eq:DD}), if the following condition is satisfied,
\begin{eqnarray}\label{eq:D2}
\frac{n}{4}\left(\sum_{\ell_i\in\Pi}\sum_{W_k\in\mc{W}}\frac{P_i}{\sigma^2_{W_k}d_{i,k}^{\alpha}}\right)^2\leq\delta,
\end{eqnarray}
then covertness is guaranteed. Equivalently, (\ref{eq:D2}) can be written as
\begin{eqnarray}\label{eq:constraint1}
\sum_{\ell_i\in\Pi}\sum_{W_k\in\mc{W}}\frac{P_i}{\sigma^2_{W_k}d_{i,k}^{\alpha}}\leq\gamma_1,
\end{eqnarray}
where $\gamma_1=2\sqrt{\frac{\delta}{n}}$.

\subsection{Maximum Throughput Covert Routing with a Single Key}
\label{sec:MT_AF}
In this section, first we find the optimal power allocation to relays of a given path $\Pi$ between Alice and Bob to maximize the throughput of covert communication over $\Pi$. Then, we design a routing algorithm that  computes the optimal path with maximum throughput from the set $\mathbf{\Pi}$ of all possible paths between Alice and Bob. While the set of all possible paths has exponential number of paths in it, we will present a routing algorithm that can find the optimal path in polynomial time.

\subsubsection{Maximum Throughput of a Given  Path}

	We consider maximizing the throughput of covert communication between Alice and Bob over a given path $\Pi$. In other words, we maximize the minimum throughput over all links in $\Pi$ such that the constraint in (\ref{eq:constraint1}) is satisfied:
	\begin{align}\label{eq:maxmin}
	\nonumber\max\left( \min_i C_i\right) ,\;i=1,\ldots,H\;\\ \text{s.t.}\;\sum_{\ell_i\in\Pi}\sum_{W_k\in\mc{W}}\frac{P_i}{\sigma^2_{W_k}d_{i,k}^{\alpha}}\leq\gamma_1,
	\end{align}
	where $C_i$ is the throughput achieved over link $\ell_i$ between relays $S(\ell_i)$ and $D(\ell_i)$, and, per Section II, we will employ 
	 \begin{eqnarray}\label{eq:c_i}
	C_i=\frac{P_i}{2\sigma_{i}^2d_{i}^{\alpha}},
	\end{eqnarray}
	where $\sigma_{i}^2$ is the variance of AWGN at $D(\ell_i)$, and  $d_i$ is the length of the link $\ell_i$.
	In the following, we show that  $\min_i C_i$  subject to the covertness constraint $\sum_{\ell_i\in\Pi}\sum_{W_k\in\mc{W}}\frac{P_i}{\sigma^2_{W_k}d_{i,k}^{\alpha}}\leq\gamma_1$ is maximized when all  links $\ell_i=\left( S(\ell_i),D(\ell_i)\right) \in\Pi$  have the same covert throughput, i.e. $C_1=\cdots=C_H$. First, let us restate (\ref{eq:maxmin}),
		\begin{align}
		\max\left( \min_i C_i\right) ,\;i=1,\ldots,H\; \text{s.t.}\;\sum_{\ell_i\in\Pi}a_iC_i\leq\gamma_1,
		\end{align}
		where,
	\begin{align}
	 a_i=\sum_{W_k\in\mc{W}}\frac{2\sigma_{i}^2d_{i}^{\alpha}}{\sigma^2_{W_k}d_{i,k}^{\alpha}}.
	\end{align}
	Suppose $C_{(1)}\leq C_{(2)}\leq \cdots \leq C_{(H)}$ are ordered $C_i$'s such that $C_{(1)}=\min_i C_i$.
	Since $ C_{(1)}\leq C_i,\;\forall\; i=1,\cdots,H$, 
	\begin{align*}
	C_{(1)}\sum_ia_i\leq \sum_iC_ia_i\leq \gamma_1,
	\end{align*} 
	and thus,
	\begin{align*}
	C_{(1)}\leq\frac{\gamma_1}{\sum_ia_i}.
	\end{align*}
		Now it remains to show that this upper-bound is achievable. This is achieved if
%	Define $\beta_{(i)}=C_{(i)}-C_{(1)}$. Thus,
%	\begin{eqnarray}
%	\gamma_1=\sum_iC_i a_i= \sum_iC_{(1)} a_i+\sum_i(C_{(i)}-C_{(1)}) g_{(i)}= C_{(1)}\sum_i a_i+\sum_i\beta_{(i)} g_{(i)}
%	\end{eqnarray}
%	 Hence,
%	\begin{eqnarray}
%	C_{(1)}=\frac{\gamma_1-\sum_i\beta_{(i)} g_{(i)}}{\sum_i a_i}.
%	\end{eqnarray}
%	Since $\sum_i\beta_{(i)} g_{(i)}\geq 0$, $C_{(1)}$ is maximized when $\sum_i\beta_{(i)} g_{(i)}= 0$ and thus,
	\begin{eqnarray}
	C_1=\cdots=C_{H}=\frac{\gamma_1}{\sum_i a_i}.
	\end{eqnarray}
	Hence, the maximum covert throughput of  a given path  in the presence of multiple Willies is given by,
	\begin{align}
	C_{\text{SK}}&=\frac{\gamma_1}{\sum_{\ell_i\in\Pi}\sum_{W_k\in\mc{W}}\frac{2\sigma_{i}^2d_{i}^{\alpha}}{\sigma^2_{W_k}d_{i,k}^{\alpha}}}\nonumber\\
	&=\frac{\sqrt{\delta}}{\sum_{\ell_i\in\Pi}\sum_{W_k\in\mc{W}}\frac{\sigma_{i}^2d_{i}^{\alpha}}{\sigma^2_{W_k}d_{i,k}^{\alpha}}}\frac{1}{\sqrt{n}}.\label{eq:rate_AF}
	\end{align}
	Using (\ref{eq:rate}), the optimal power that a relay $S(\ell_i)\in\Pi$ should transmit with to obtain the maximum covert throughput in (\ref{eq:rate_AF}) is,
	\begin{align*}
	P_i&=2\sigma_{i}^2d_i^{\alpha}C_{\text{SK}}\\
	&=\frac{2\sqrt{\delta}\sigma_{i}^2d_i^{\alpha}}{\sum_{\ell_j\in\Pi}\sum_{W_k\in\mc{W}}\frac{\sigma_{j}^2d_{j}^{\alpha}}{\sigma^2_{W_k}d_{j,k}^{\alpha}}} \hspace{-2pt}\frac{1}{\sqrt{n}}.
	\end{align*}
	
	\subsubsection{MT-SK Routing Algorithm}\label{sec:MT-SK}
	
	In this section, we find the optimal path  with maximum throughput  between Alice and Bob. From (\ref{eq:rate_AF}),  the path that maximizes the covert throughput  is the path that  minimizes $\sum_{\ell_i\in\Pi}\sum_{W_k\in\mc{W}}\frac{\sigma_{i}^2d_{i}^{\alpha}}{\sigma^2_{W_k}d_{i,k}^{\alpha}}$. %Hence, in order to find the path with maximum throughput,  we should find the path for which $\sum_{\ell_i\in\Pi}\sum_{W_k\in\mc{W}}\frac{\sigma_{i}^2d_{i}^{\alpha}}{\sigma^2_{W_k}d_{i,k}^{\alpha}}$ is minimum.
	Let us define the ``cost" of a maximum covert throughput path $\Pi$ with a single key as, 
	\begin{eqnarray}
	{\omega}_{\text{MT-SK}}(\Pi)=\sum_{\ell_i\in\Pi}\sum_{W_k\in\mc{W}}\frac{\sigma_{i}^2d_{i}^{\alpha}}{\sigma^2_{W_k}d_{i,k}^{\alpha}},
	\end{eqnarray}
	and the cost of communication over  a  link  $\ell_i=\left( S(\ell_i),D(\ell_i)\right) $ as,
	\begin{eqnarray}
	\omega_{\text{MT-SK}}(\ell_i)=\sum_{W_k\in\mc{W}}\frac{\sigma_{i}^2d_{i}^{\alpha}}{\sigma^2_{W_k}d_{i,k}^{\alpha}}.
	\end{eqnarray} 
	%The minimum cost (maximum throughput) path is obtained by solving a shortest path problem with link costs $\omega(\ell_i)$.
	Since the cost of each link does not depend on other links, we can obtain the  minimum cost (maximum throughput) path by assigning the cost $\omega_{\text{MT-SK}}(\ell_i)$ to every potential link $\ell_i$ of the network, and solving a shortest-path problem.
	There are several classical shortest path algorithms that can be used for this purpose. In this paper, we use Dijkstra's algorithm.
	
\subsection{Minimum Delay  Covert Routing with a Single Key}
\label{sec:MD_AF}

In this section, first we  find the optimal power allocation to minimize the end-to-end delay of a given path when transmitting a message from Alice to Bob in the presence of multiple Willies. Then we design a routing algorithm to choose the path with minimum end-to-end  delay  between Alice and Bob.

\subsubsection{Minimum Delay of a Given  Path}\label{sec:MD_CR_AFR}

Suppose we have a multi-hop path from Alice to Bob. Here our goal is to minimize the end-to-end delay of transmitting a message covertly over a given path $\Pi$ from Alice to Bob,
  such that the constraint in (\ref{eq:constraint1}) is satisfied. We define the average delay of the $i^{th}$ link, denoted by $\Delta_i$, as the inverse of the link covert throughput,  $\Delta_i=\frac{1}{C_i}$, and thus the end-to-end delay can be written  as,
  \[\Delta_{\text{SK}}(\Pi)=\sum_{\ell_i\in\Pi}\Delta_i=\sum_{\ell_i\in\Pi}\frac{1}{C_i}.\]
  Hence, we want to solve the following problem,
\begin{align}\label{eq:ar_mindelay}
\min \Delta_{\text{SK}}(\Pi),\quad \text{s.t.}\quad\sum_{\ell_i\in\Pi}\sum_{W_k\in\mc{W}}\frac{P_i}{\sigma^2_{W_k}d_{i,k}^{\alpha}}\leq\gamma_1.
\end{align}
From (\ref{eq:rate}),
\begin{eqnarray}
\Delta_i=\frac{1}{C_i}=\frac{2\sigma_{i}^2d_{i}^{\alpha}}{P_i}.
\end{eqnarray}
%Hence, (\ref{eq:ar_mindelay}) can be written as,
%\begin{align}\label{eq:ar_mindelay2}
%\min\sum_{i=1}^H\frac{2\sigma_{i}^2d_{i}^{\alpha}}{P_i},\; \text{s.t.}\quad\sum_{\ell_i\in\Pi}\sum_{W_k\in\mc{W}}\frac{P_i}{\sigma^2_{W_k}d_{i,k}^{\alpha}}\leq\gamma_1.
%\end{align}
Let us define,
\begin{align}\label{eq:af_Xg}
 b_i=\sum_{W_k\in\mc{W}}\frac{2\sigma_{i}^2d_{i}^{\alpha}}{\sigma^2_{W_k}d_{i,k}^{\alpha}}.
\end{align}
Substituting  $b_i$ in (\ref{eq:ar_mindelay}), our optimization problem is,
\begin{align}\label{eq:ar_mindelay3}
\min\Delta_{\text{SK}}(\Pi),\; \text{s.t.}\quad \sum_{\ell_i\in\Pi}\frac{b_i}{\Delta_i}\leq\gamma_1.
\end{align}
In (\ref{eq:ar_mindelay3}) the optimization objective is linear and the constraint   is a convex set, and thus (\ref{eq:ar_mindelay3}) is a convex optimization problem. Hence, %a set $\{\Delta_1,\ldots,\Delta_H\}$
a point that minimizes $\Delta_{\text{SK}}(\Pi)$ in (\ref{eq:ar_mindelay3}) is a global minimum.  Since left side of the constraint in (\ref{eq:ar_mindelay3}) is a decreasing function of $\Delta_i$ and our goal is to minimize $\sum_{\ell_i\in\Pi}\Delta_i$, the constraint is active and becomes
\begin{align}\label{eq:constr1}
\sum_{\ell_i\in\Pi}\frac{b_i}{\Delta_i}=\gamma_1.
\end{align}
In order to solve this optimization problem, we use the Lagrange multipliers technique. Thus, we should solve the following Lagrangian equations and the constraint  (\ref{eq:constr1}) simultaneously,
\begin{align*}
\frac{\partial}{\partial \Delta_i}\left\lbrace \sum_{j=1}^H\Delta_j+\lambda\left(  \sum_{j=1}^H\frac{b_j}{\Delta_j}-\gamma_1\right) \right\rbrace=0,\; i=1,\ldots, H.
\end{align*}
Taking the derivatives of the Lagrangian functions the following equations are obtained,
\begin{align}\label{eq:lagrange1}
1-\lambda\frac{b_i}{\Delta_i^2}=0,\; i=1,\ldots, H.
\end{align}
Substituting $\Delta_i$ from (\ref{eq:lagrange1}) into (\ref{eq:constr1}), we obtain,
\begin{align}\label{eq:lambda}
\lambda=\frac{1}{\gamma_1^2}\left( \sum_i\sqrt{b_i}\right) ^2
\end{align}
Hence, after substituting $\lambda$ from (\ref{eq:lambda}) into (\ref{eq:lagrange1}), $\Delta_i$ is given by,
\begin{align}\label{eq:delta_i}
\Delta_i=\frac{1}{\gamma_1}\sqrt{b_i}\sum_{j=1}^H\sqrt{b_j}.
\end{align}
Therefore, we have,
\begin{align}\label{eq:af_delayX}
\sum_{i=1}^H\Delta_i=\frac{1}{\gamma_1}\left( \sum_{j=1}^H\sqrt{b_j}\right) ^2.
\end{align}
From  (\ref{eq:af_Xg}), (\ref{eq:ar_mindelay3}) and (\ref{eq:af_delayX}), the minimum end-to-end delay over a given path $\Pi$ can be written as,
\begin{align}\label{eq:af_delay2}
\sum_{i=1}^H\Delta_i=\frac{2}{\gamma_1}\left( \sum_{\ell_i\in\Pi}\sqrt{\sum_{W_k\in\mc{W}}\frac{\sigma_{i}^2d_{i}^{\alpha}}{\sigma^2_{W_k}d_{i,k}^{\alpha}}}\right) ^2.
\end{align}
In order to obtain the minimum delay,  relay $S(\ell_i)$ along the path $\Pi$ should transmit to relay $D(\ell_i)$ with power,
\begin{align}
P_i=\frac{2\sigma_{i}^2d_{i}^{\alpha}}{\Delta_i},
\end{align}
where from (\ref{eq:af_Xg}) and (\ref{eq:delta_i}), 
\begin{align}
\Delta_i\hspace{-2pt}=\hspace{-3pt}\left(\hspace{-2pt} \frac{1}{\sqrt{\delta}}\sqrt{\sum_{W_k\in\mc{W}}\frac{\sigma_{i}^2d_{i}^{\alpha}}{\sigma^2_{W_k}d_{i,k}^{\alpha}}}\sum_{\ell_j\in\Pi}\hspace{-4pt} \sqrt{\sum_{W_k\in\mc{W}}\frac{\sigma_{j}^2d_{j}^{\alpha}}{\sigma^2_{W_k}d_{j,k}^{\alpha}}}\right)\hspace{-4pt} \sqrt{n}.
\end{align}
\subsubsection{MD-SK Routing Algorithm}\label{sec:MD-SK}

In this section, our goal is to find the path $\Pi$ with minimum end-to-end delay  from Alice to Bob. From (\ref{eq:af_delay2}),  in order to find the path with minimum delay, we should find a path for which
\begin{align*}\label{eq:delay2}
\Delta_{\text{SK}}(\Pi)=\sum_{\ell_i\in\Pi}\Delta_i=\frac{2}{\gamma_1}\left( \sum_{\ell_i\in\Pi}\sqrt{\sum_{W_k\in\mc{W}}\frac{\sigma_{i}^2d_{i}^{\alpha}}{\sigma^2_{W_k}d_{i,k}^{\alpha}}}\right) ^2,
\end{align*} 
 is minimum.
Let us define the cost of covert communication to minimize the end-to-end delay with a single key (MD-SK) of a path $\Pi$ as,
\begin{eqnarray}
{\omega}_{\text{MD-SK}}(\Pi)= \sum_{\ell_i\in\Pi}\sqrt{\sum_{W_k\in\mc{W}}\frac{\sigma_{i}^2d_{i}^{\alpha}}{\sigma^2_{W_k}d_{i,k}^{\alpha}}}
\end{eqnarray}
This can be attained by assigning the link cost:
\begin{eqnarray}\label{eq:cost_AF_delay}
\omega_{\text{MD-SK}}(\ell_i)=\sqrt{\sum_{W_k\in\mc{W}}\frac{\sigma_{i}^2d_{i}^{\alpha}}{\sigma^2_{W_k}d_{i,k}^{\alpha}}}
\end{eqnarray} 
to every potential link $\ell_i$ in the network. 
Clearly, a path $\Pi$ that minimizes ${\omega}_{\text{MD-SK}}(\Pi)= \sum_{\ell_i\in\Pi}\omega_{\text{MD-SK}}(\ell_i)$ also minimizes the end-to-end delay $\Delta_{\text{SK}}(\Pi)$. 
Hence, the problem is reduced to a  shortest path problem  with link costs  $\omega_{\text{MD-SK}}(\ell_i)$ given by (\ref{eq:cost_AF_delay}).

\section{Covert Communication with Independent Keys at the Relays (IK)}
\label{sec:DF}
In this section, we consider  multi-hop covert communication between Alice and Bob in the presence of multiple collaborating Willies with independent keys at the relays. 
In this approach, each relay along the path between Alice and Bob re-encodes the message with a different key, and then forwards it to the next relay until it is delivered to the destination, Bob.
Since the message is encoded with different and independent keys at each hop,  unlike the previous approach the signal sent over each hop  is independent of the signal sent over  other hops.

\subsection{Covertness Analysis of Covert Communication with Independent Keys}\label{sec:cov_df}

		Suppose the message is sent over a path $\Pi$ from Alice to Bob, and  $\mathbb{Q}_0$ is the joint probability distribution   of Willies' observations over  all hops and  $n$ channel uses under hypothesis $H_0$,
		and  $\mathbb{Q}_1$ is the joint probability distribution  of Willie's observations   over all hops and $n$ channel uses under hypothesis $H_1$. Since the message is encoded with different independent key sequences at each hop, the codewords sent over  different  hops are independent. Also, the noise is AWGN and thus is independent across different hops.
		Hence,
		\begin{align}
		\mb{Q}_0=\prod_{\ell_i\in \Pi} \mb{Q}_0^i ,\;\text{and,}\;\mb{Q}_1=\prod_{\ell_i\in \Pi}\mb{Q}_1^i
		\end{align} 
		where $\mb{Q}_0^i$ and $\mb{Q}_1^i$ are the joint probability distributions of Willies' observations   over the $i^{th}$ hop under hypotheses $H_0$ and $H_1$, respectively.
		Suppose Willies apply the optimal hypothesis test to make a decision on the end-to-end communication.
		From the independence of the observations across hops, the end-to-end relative entropy between $\mb{Q}_1$ and $\mb{Q}_0$ is,
		\begin{align}\label{eq:DQ_sum1}
		\mb{D}(\mb{Q}_1\|\mb{Q}_0)&=\sum_{\ell_i\in \Pi} \mb{D}(\mb{Q}_1^i\|\mb{Q}_0^i).
		\end{align}
		At each hop, Willies combine their observations across Willies and across time. 
		Using the same approach as in Section \ref{sec:covertness_AF}, the relative entropy between $\mb{Q}_1^i$ and $\mb{Q}_0^i$  is,
		\begin{align}\label{eq:D_one}
		&\nonumber\mathbb{D}(\mathbb{Q}_1^i\|\mathbb{Q}_0^i)\\
		&=\frac{n}{2}\left(\sum_{W_k\in\mc{W}} \frac{P_i} {\sigma^2_{W_k}d_{i,k}^{\alpha}}-\ln\left(1+\sum_{W_k\in\mc{W}}\frac{P_i}{\sigma^2_{W_k}d_{i,k}^{\alpha}}\right)\right)
		\end{align} 
		Using the fact that $\ln(1+x)\geq x-\frac{x^2}{2}$ for $x\geq 0$,
		\begin{eqnarray}\label{eq:D4}
		\mathbb{D}(\mathbb{Q}_1^i\|\mathbb{Q}_0^i)\leq \frac{n}{4}\left(\sum_{W_k\in\mc{W}}\frac{P_i}{\sigma^2_{W_k}d_{i,k}^{\alpha}}\right)^2.
		\end{eqnarray}
		Hence, from (\ref{eq:DQ_sum1}) and (\ref{eq:D4}) the end-to-end relative entropy between $\mb{Q}_1$ and $\mb{Q}_0$ can be written as,
	\begin{align}\label{eq:DQ_sum}
	\mb{D}(\mb{Q}_1\|\mb{Q}_0)&=\sum_{\ell_i\in \Pi} \mb{D}(\mb{Q}_1^i\|\mb{Q}_0^i)\nonumber\\
	&\leq\frac{n}{4}\sum_{\ell_i\in\Pi}\left(\sum_{W_k\in\mc{W}}\frac{P_i}{\sigma^2_{W_k}d_{i,k}^{\alpha}}\right)^2.
	\end{align}
		Combining (\ref{eq:covert3}) and (\ref{eq:DQ_sum}), in order to guarantee the end-to-end covertness it suffices to have,
		\begin{eqnarray}\label{eq:df_cov_cons}
		\frac{n}{4}\sum_{\ell_i\in\Pi}\left(\sum_{W_k\in\mc{W}}\frac{P_i}{\sigma^2_{W_k}d_{i,k}^{\alpha}}\right)^2\leq\delta.
		\end{eqnarray}
		Setting $\gamma_2={\frac{4\delta}{n}}$ the covertness constraint (\ref{eq:DQ_sum}) can be written as,
		\begin{eqnarray}\label{eq:constraint3}
	\sum_{\ell_i\in\Pi}\left(\sum_{W_k\in\mc{W}}\frac{P_i}{\sigma^2_{W_k}d_{i,k}^{\alpha}}\right)^2\leq\gamma_2.
		\end{eqnarray}
		
		\subsection{Comparison of Multi-Hop Covert Communication with a Single Key and with Independent Keys}\label{sec:comp}
		
		Consider the covertness constraint of covert communication with a single key  (\ref{eq:D2}) in Section \ref{sec:cov_af} and let,
		\begin{eqnarray}\label{eq:af_D22}
		B_{\text{SK}}=\frac{n}{4}\left(\sum_{\ell_i\in\Pi}\sum_{W_k\in\mc{W}}\frac{P_i}{\sigma^2_{W_k}d_{i,k}^{\alpha}}\right)^2.
		\end{eqnarray}
		 Also, consider the covertness constraint of covert communication with independent keys at the relays (\ref{eq:df_cov_cons}) in Section \ref{sec:cov_df} and let, 
			\begin{eqnarray}\label{eq:df_cov_cons}
		B_{\text{IK}}=	\frac{n}{4}\sum_{\ell_i\in\Pi}\left(\sum_{W_k\in\mc{W}}\frac{P_i}{\sigma^2_{W_k}d_{i,k}^{\alpha}}\right)^2.
			\end{eqnarray}
			Clearly, for the  same path $\Pi$ and same powers $P_i$, $B_{\text{IK}}<B_{\text{SK}}$. 
			Hence, with the same covertness constraint $\delta$,  communication with  independent keys approach  compared to communication with a single key approach  allows higher powers while maintaining covertness, which results in higher throughput and lower delay.  Hence, we expect  covert communication with independent keys  to have better performance than covert communication with a single key. We will show this in more detail with simulations for various parameters of the network in Section \ref{sec:numerical}. Note that the better performance of the scheme with independent keys comes at the expense of more key bits.
		
		\subsection{Maximum Throughput Covert Communication with Independent Keys}
		\label{sec:MT_DF}
		
		In this section, we characterize the optimum power allocation to the relays along a given path $\Pi$ from Alice to Bob to maximize the covert throughput when using independent keys at relays.  Also, we design a routing algorithm that computes the maximum throughput path.% $\Pi$ from the set $\mathbf{\Pi}$ of all potential paths between Alice and Bob.
		
		\subsubsection{Maximum Throughput of a Given  Path}
		
		Here we find the optimal power allocation  on a  given path $\Pi$ between Alice and Bob to maximize the covert throughput. In order to find the maximum covert throughput, we should maximize the minimum throughput over all hops in $\Pi$   such that the constraint in (\ref{eq:constraint3}) is satisfied,
		\begin{align}\label{eq:maxmin3}
	\nonumber	\max\left( \min_i C_i\right) ,\;i=1,\ldots,H\\
		 \text{s.t.}\;\sum_{\ell_i\in\Pi}\left(\sum_{W_k\in\mc{W}} \frac{P_i}{\sigma^2_{W_k}d_{i,k}^{\alpha}}\right)^2\leq\gamma_2.
		\end{align}
		This optimization is similar to the optimization in Section \ref{sec:MT_AF}, and can be solved in the same way. The detailed solution is presented in Appendix \ref{ap:2}.
	Hence, the maximum covert throughput of a given path   with independent keys at the relays is,
	\begin{align}\nonumber
	C_{\text{IK}}&=\frac{\sqrt{\gamma_2}}{\sqrt{\sum_{\ell_i\in\Pi} \left(\sum_{W_k\in\mc{W}} \frac{2\sigma_{i}^2d_{i}^{\alpha}}{\sigma^2_{W_k}d_{i,k}^{\alpha}}\right) ^2 }}\\
	&=\frac{\sqrt{\delta}}{\sqrt{\sum_{\ell_i\in\Pi} \left(\sum_{W_k\in\mc{W}}\label{maximum_rate_DF} \frac{\sigma_{i}^2d_{i}^{\alpha}}{\sigma^2_{W_k}d_{i,k}^{\alpha}}\right) ^2 }}\frac{1}{\sqrt{n}}.
	\end{align} 
	Each relay along the path $\Pi$ should transmit its message to the next relay with optimal power,
	\begin{align}\nonumber
	P_i&=2\sigma_{i}^2d_{i}^{\alpha} C_{\text{IK}}\\
	&=\frac{2\sigma_{i}^2d_{i}^{\alpha}\sqrt{\delta}}{\sqrt{\sum_{\ell_j\in\Pi} \left(\sum_{W_k\in\mc{W}} \frac{\sigma_{j}^2d_{j}^{\alpha}}{\sigma^2_{W_k}d_{j,k}^{\alpha}}\right) ^2 }}\frac{1}{\sqrt{n}}.
	\end{align} 
	
\subsubsection{MT-IK Routing Algorithm}\label{sec:MT-IK}

 From (\ref{maximum_rate_DF})  in order to find the maximum throughput path $\Pi$ between Alice and Bob, we should find the path $\Pi$ for which ${\sum_{\ell_i\in\Pi} \left( \sum_{W_k\in\mc{W}}\frac{2\sigma_{i}^2d_{i}^{\alpha}}{\sigma^2_{W_k}d_{i,k}^{\alpha}}\right) ^2 }$ is minimum.
Hence, define the cost of covert communication to maximize the throughput with independent keys at the relays (MT-IK)  of a path $\Pi$ as,
\begin{eqnarray}\label{eq:cost_C_DF}
{\omega}_{\text{MT-IK}}(\Pi)=\sum_{\ell_i\in\Pi} \left(\sum_{W_k\in\mc{W}} \frac{\sigma_{i}^2d_{i}^{\alpha}}{\sigma^2_{W_k}d_{i,k}^{\alpha}}\right) ^2.
\end{eqnarray}
%The minimum cost  path is the maximum covert throughput path. 
Assign the following link cost $\omega({\ell_i})$ to every potential link in the network:
\begin{eqnarray}
\omega_{\text{MT-IK}}(\ell_i)=\left( \sum_{W_k\in\mc{W}}\frac{\sigma_{i}^2d_{i}^{\alpha}}{\sigma^2_{W_k}d_{i,k}^{\alpha}}\right) ^2
\end{eqnarray} 
 and find the shortest path $\Pi$ with link costs $\omega_{\text{MT-IK}}(\ell_i)$ using any shortest path routing algorithm.  
  
\subsection{Minimum Delay  Covert Routing with Independent Keys}
In this section, first we find the suitable power allocation to  minimize the end-to-end delay of covert communication over a given path. Next, we propose a routing algorithm to find the minimum end-to-end delay path from the set of all paths $\mathbf{\Pi}$ between Alice and Bob.
\label{sec:MD_DF}
\subsubsection{Minimum Delay of a Given  Path}

Here the goal is to minimize the end-to-end delay of a given path $\Pi$ such that the constraint in (\ref{eq:constraint3}) is satisfied,
\begin{align}\label{eq:df_mindelay}
\min\:\Delta_{\text{IK}}(\Pi),\quad \text{s.t.}\;\sum_{\ell_i\in\Pi}\left(\sum_{W_k\in\mc{W}} \frac{P_i}{\sigma^2_{W_k}d_{i,k}^{\alpha}}\right) ^2\leq\gamma_2.
\end{align}
Define,
\begin{align}
h_i=\left(\sum_{W_k\in\mc{W}} \frac{2\sigma_{i}^2d_{i}^{\alpha}}{\sigma^2_{W_k}d_{i,k}^{\alpha}}\right) ^2.
\end{align}
Substituting $\Delta_{\text{IK}}(\Pi)=\sum_{i=1}^H\Delta_i$ and $h_i$ in (\ref{eq:df_mindelay}), our optimization problem is,
\begin{align}\label{eq:df_mindelay3}
\min\sum_{i=1}^H\Delta_i,\quad\text{s.t.}\quad \sum_{\ell_i\in\Pi}\frac{h_i}{\Delta_i^2}\leq\gamma_2.
\end{align}
The objective function in (\ref{eq:df_mindelay3}) is linear and the constraint is a convex set. Hence, (\ref{eq:df_mindelay3}) is a convex optimization problem and any point that minimizes the objective function is a global minimum as well. 
Using the same reasoning as in Section \ref{sec:MD_CR_AFR}, the constraint in (\ref{eq:df_mindelay3}) is active and thus the inequality
constraint in (\ref{eq:df_mindelay3}) can be substituted by the following equality constraint, 
\begin{align}\label{eq:df_constr2}
\sum_{\ell_i\in\Pi}\frac{h_i}{\Delta_i^2}=\gamma_2.
\end{align}
In order to solve this optimization problem, we use the Lagrange multipliers technique. Thus, we should solve the following equations and the constraint (\ref{eq:df_constr2}) simultaneously,
\begin{align*}
&\frac{\partial}{\partial \Delta_i}\left\lbrace \sum_{j=1}^H\Delta_j+\lambda\left(  \sum_{j=1}^H\frac{h_j}{\Delta_j^2}-\gamma_2\right) \right\rbrace=0,\\
&\qquad\qquad\qquad\qquad\qquad\qquad\qquad i=1,\ldots, H.
\end{align*}
Setting  the derivatives to zero, we have,
\begin{align*}
1-2\lambda\frac{h_i}{\Delta_i^3}=0,\; i=1,\ldots, H,
\end{align*}
and thus,
\begin{align}\label{eq:df_lagrange1}
\Delta_i=\left( 2\lambda h_i\right)^{1/3},\; i=1,\ldots, H.
\end{align}
Substituting $\Delta_i$ from (\ref{eq:df_lagrange1}) into (\ref{eq:df_constr2}),
\begin{align}\label{eq:df_lambda}
\lambda=\frac{1}{(2\gamma_2)^{3/2}}\left( \sum_i{h_i}^{1/3}\right) ^{3/2}.
\end{align}
Hence, by substituting $\lambda$ from (\ref{eq:df_lambda}) into (\ref{eq:df_lagrange1}) we have,
\begin{align}
\Delta_i=\frac{1}{\sqrt{\gamma_2}}{h_i}^{1/3}\left( \sum_{j=1}^H{h_j}^{1/3}\right) ^{1/2}.
\end{align}
Thus, the minimum end-to-end delay of sending a message covertly from Alice to Bob over a given path $\Pi$ is,
\begin{align*}
\Delta_{\text{IK}}(\Pi)&=\sum_{i=1}^H\Delta_i\\
&=\frac{1}{\sqrt{\gamma_2}}\sum_{i=1}^H{h_i}^{1/3}\left( \sum_{i=1}^H{h_i}^{1/3}\right) ^{1/2}\\
&=\frac{1}{\sqrt{\gamma_2}}\left( \sum_{i=1}^H{h_i}^{1/3}\right) ^{3/2}\\
&=\frac{2}{\sqrt{\gamma_2}}\left( \sum_{\ell_i\in\Pi}{\left(\sum_{W_k\in\mc{W}} \frac{\sigma_{i}^2d_{i}^{\alpha}}{\sigma^2_{W_k}d_{i,k}^{\alpha}}\right) }^{1/3}\right) ^{3/2}.\label{eq:df_delay1}\numberthis
\end{align*}
In order to attain the minimum end-to-end delay,  relay $S(\ell_i)$ should transmit with power,
\begin{align}
P_i=\frac{2\sigma_{i}^2d_{i}^{\alpha}}{\Delta_i}
\end{align}
where, 
\begin{align}
\Delta_i\hspace{-2pt}=\hspace{-2pt}\frac{1}{\sqrt{\gamma_2}}\hspace{-2pt}\left( {\sum_{W_k\in\mc{W}}\frac{2\sigma_{i}^2d_{i}^{\alpha}}{\sigma^2_{W_k}d_{i,k}^{\alpha}}}\right)^{\hspace{-3pt}1/3}\hspace{-4pt}\left( \sum_{\ell_j\in\Pi}\hspace{-2pt}\left( {\sum_{W_k\in\mc{W}}\frac{2\sigma_{j}^2d_{j}^{\alpha}}{\sigma^2_{W_k}d_{j,k}^{\alpha}}}\right)^{\hspace{-4pt}1/3}\hspace{-1pt}\right)^{\hspace{-4pt}1/2}\hspace{-3pt} .
\end{align}
 
 \begin{figure}
 	\centering
 	\includegraphics[width=.5\textwidth]{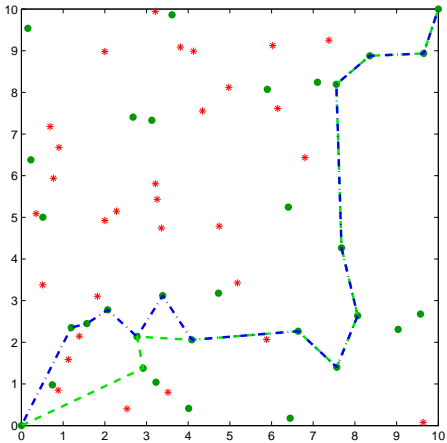}
 	\caption{A snapshot of the network when 30 system nodes (green circles) and 30 Willies (red stars) are present in the network. All nodes are distributed uniformly at random over the network. The covertness factor is set to $\delta=0.05$, and the path-loss exponent to $\alpha=3$. The path that achieves the maximum throughput with the MT-SK algorithm is shown by blue dash-dot lines, and the path that achieves the maximum throughput with the MT-IK algorithm is shown by green dashed lines. The maximum covert throughput using the MT-SK algorithm is $0.0024/\sqrt{n} $ and the maximum covert throughput using the MT-IK algorithm is $0.0056/\sqrt{n}$.}
 	\label{fig:snapshot_rate}
 \end{figure}
\subsubsection{MD-IK Routing Algorithm}\label{sec:MD-IK}

In order to compute the path with maximum throughput $\Pi$  between Alice and Bob, we should  find the path  for which the end-to-end delay, 
\begin{align}
\Delta_{\text{IK}}(\Pi)=\frac{2}{\sqrt{\gamma_2}}\left( \sum_{\ell_i\in \Pi}{\left(\sum_{W_k\in\mc{W}} \frac{\sigma_{i}^2d_{i}^{\alpha}}{\sigma^2_{W_k}d_{i,k}^{\alpha}}\right) }^{1/3}\right) ^{3/2},
\end{align}
is minimized.
Define the cost of covert communication with minimum end-to-end delay using independent keys at the relays (MD-IK) over  a path $\Pi$ as,
\begin{eqnarray}
{\omega}_{\text{MD-IK}}(\Pi)=\sum_{\ell_i\in\Pi} \left( \sum_{W_k\in\mc{W}}\frac{\sigma_{i}^2d_{i}^{\alpha}}{\sigma^2_{W_k}d_{i,k}^{\alpha}}\right) ^{1/3}.
\end{eqnarray}
Assign the link cost $\omega_{\text{MD-IK}}({\ell_i})$ to every link $\ell_i$ in the network,
\begin{eqnarray}
\omega_{\text{MD-IK}}(\ell_i)=\left( \sum_{W_k\in\mc{W}}\frac{\sigma_{i}^2d_{i}^{\alpha}}{\sigma^2_{W_k}d_{i,k}^{\alpha}}\right) ^{1/3},
\end{eqnarray} 
 and apply any  shortest-path algorithm  to find the path with minimum cost from Alice to Bob, which is the desired path $\Pi^*$.
 
 \begin{figure}
 	\centering
 	\includegraphics[width=.5\textwidth]{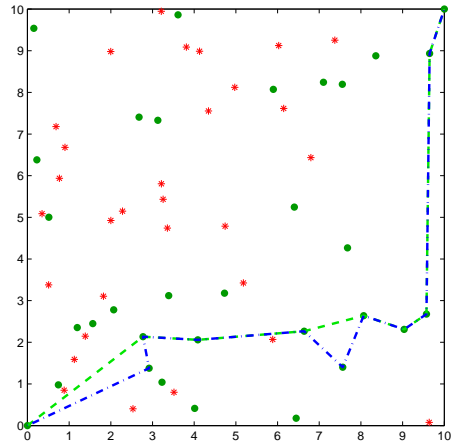}
 	\caption{A snapshot of the network when 30 system nodes (green circles) and 30 Willies (red stars) are present in the network. All nodes are distributed uniformly at random over the network. The covertness factor is set to $\delta=0.05$, and the path-loss exponent to $\alpha=3$. The path achieving the minimum delay with the MD-SK algorithm is shown by blue dash-dotted lines, and the path achieving the minimum delay with the MD-IK algorithm is shown by green dashed lines. The minimum end-to-end delay using the MD-SK algorithm is $2448.8\sqrt{n} $ and the minimum end-to-end delay using the MD-IK algorithm is $1048.1\sqrt{n} $.}
 	\label{fig:snapshot_delay}
 \end{figure}
 
\section{Numerical Results}\label{sec:numerical}

%In this section, we show how using  multi-hop communication with optimal power allocation over a given path  improves the performance of  one-hop communication from Alice to Bob.   Further,  we will show that how using the proposed routing algorithms improve the performance of multi-hop communication over a given path. 
In this section, we evaluate and compare the performance of the  routing algorithms  proposed in this paper   numerically.
A wireless network on a $d\times d$ square on the 2-D plane with corners $(0,0), (0,d), (d,0),(d,d)$ is considered. 
In all simulations, Alice (source) is located at point $(0,0)$ and Bob (destination) is located at point  $(d,d)$.
Multiple friendly system nodes and multiple Willies are distributed uniformly at random over the network.
\begin{figure}
	\centering
	\includegraphics[width=.5\textwidth]{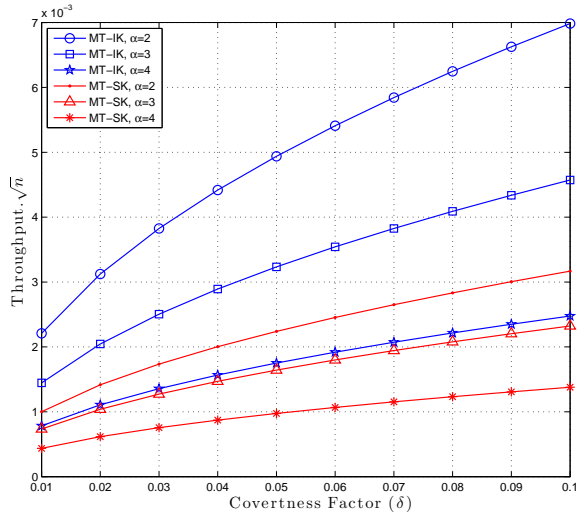}
	\caption{Maximum throughput versus covertness factor $\delta$, where 30 system nodes and 30 Willies are present in the network.}
	\label{fig:rate_vs_delta}
\end{figure}
We consider fully connected networks. For all routing algorithms, we assign the link costs described in 
% Sections \ref{sec:MT-SK}, \ref{sec:MD-SK}, \ref{sec:MT-IK}, and \ref{sec:MD-IK}  
previous sections to every link in the network,   and then apply the Dijkstra's algorithm to find the shortest (minimum cost) path from Alice to Bob in each case.
Since Dijkstra's algorithm is a polynomial time algorithm,  the computational complexity of the proposed algorithms 
is also polynomial in the size of the network, and hence the proposed routing algorithms are efficient.
%Note that since the costs of communication in all algorithms developed in this work are  
%In each case, to consider the effect of path-loss exponent on the algorithms we present  results for  $\alpha=2,3, \;\text{and}\; 4$.

Figs. \ref{fig:snapshot_rate} and \ref{fig:snapshot_delay} show one snapshot of the network when 30 system nodes and 30 Willies are present. In both figures we set the  path-loss exponent to $\alpha=3$ and the covertness factor to $\delta=0.05$. In Fig. \ref{fig:snapshot_rate},  the  maximum throughput paths obtained by the MT-SK and MT-IK algorithms are shown. In this case, the covert throughput of one-hop communication from Alice to Bob  with covertness factor of $\delta=0.05$ is $8.4476\times10^{-5}/\sqrt{n}$.
When using multi-hop communication, the maximum covert throughput of the MT-SK algorithm is $0.0024/\sqrt{n} $ and the maximum covert throughput of the MT-IK algorithm is $0.0056/\sqrt{n}$. 
Thus, both MT-SK and MT-IK algorithms improve the performance of one hop covert communication significantly, since
they both choose paths that avoid Willies, and allocate the covertness factor to the links on each path  to increase the covert throughput. As expected from  Section \ref{sec:comp}, MT-IK offers a higher covert throughput compared to MT-SK.

In Fig. \ref{fig:snapshot_delay}, the  minimum delay paths selected  by the MD-SK and  MD-IK algorithms are shown. 
The minimum delay of one-hop covert communication from Alice to Bob, which is defined as the inverse of the covert throughput of the link from Alice to Bob, is  $11838\sqrt{n}$. 
The minimum end-to-end delay of MD-SK  is $2448.8\sqrt{n} $, and the minimum end-to-end delay of MD-IK  is $1048.1\sqrt{n} $.
Thus, both the MD-SK and  MD-IK algorithms improve the performance of one-hop covert communication significantly.  Again, both paths  avoid Willies by taking detours. 
Because of the different allocation of the covertness factors   to the links along each path,  the optimal paths and the optimal end-to-end delays are different, and, as expected, the MD-IK algorithm offers a smaller end-to-end delay than the MD-SK algorithm.

In the remainder of this section, we consider the effect of different parameters of the network on the performance of the MT-SK,   MT-IK, MD-SK, and MD-IK algorithms.
We average our results over 100 randomly generated realizations of the  network with different seeds, and  with uniform distribution of  system nodes and  Willies.
Our performance metric is the average throughput over different realizations of the network for the MT-SK and MT-IK algorithms, and the average  end-to-end delay over different realizations of the network for the MD-SK and MD-IK algorithms.
In order to have  precise comparisons, we use the same placements of the system nodes and Willies  in different cases. 

\begin{figure}
	\centering
	\includegraphics[width=.5\textwidth]{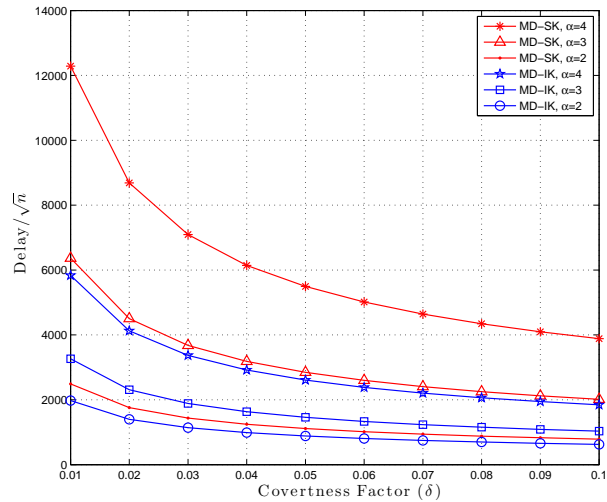}
	\caption{End-to-end delay versus covertness factor $\delta$, where 30 system nodes and 30 Willies are present in the network.}
	\label{fig:delay_vs_delta}
\end{figure}
\textbf{Effect of the covertness factor $\delta$.}
Fig. \ref{fig:rate_vs_delta} shows the maximum throughput of MT-SK and MT-IK versus the covertness factor $\delta$, when 30 system nodes and 30 Willies are present in the network, and $\delta$ changes from 0.01 to 0.1.
As can be seen, the performance of MT-IK for different $\delta$s and for different path-loss exponents $\alpha$ is better than MT-SK. Also, as expected, as $\delta$ increases, higher covert throughputs can be achieved.
% because when  the covertness constraint become looser the higher throughputs can be achieved.% The plots of throughputs versus $\delta$ resemble the plot of a  of $\sqrt{\delta}$ 

In Fig. \ref{fig:delay_vs_delta}, the minimum end-to-end delay of MD-SK and MD-IK versus  the covertness factor $\delta$ is shown, where 30 system nodes and 30 Willies are present in the network, and $\delta$ changes from 0.01 to 0.1. It is apparent that using the MD-IK algorithm, a smaller end-to-end delay can be achieved. As $\delta$ increases, each relay can transmit with higher power (higher throughput), and thus the end-to-end delay decreases for both MD-SK and MD-IK.

\begin{figure}
	\centering
	\includegraphics[width=.5\textwidth]{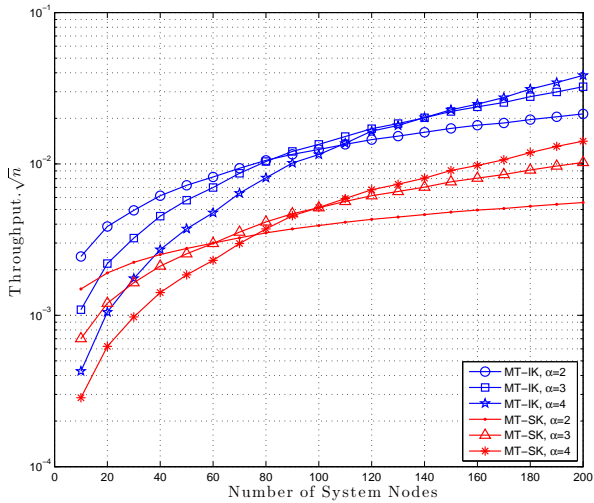}
	\caption{Throughput versus the number of system nodes, when  30 Willies are present in the network and the covertness factor $\delta=0.05$.}
	\label{fig:rate_vs_num_system}
\end{figure}

\textbf{Effect of the number of system nodes.}
In Fig. \ref{fig:rate_vs_num_system}, the maximum covert throughput of MT-SK and MT-IK versus the number of system nodes, when $\delta=0.05$ and 30 Willies are present in the network, are shown. As the number of system nodes increases, the maximum throughputs that  MT-SK and MT-IK can achieve increase, since the path can take more detours to avoid  Willies. The maximum covert throughput of  MT-IK is always larger than the maximum covert throughput of MT-SK. 
Further, as the number of system nodes increases, the maximum throughput of  MT-SK increases slowly. 
The reason is that for MT-SK, when the number of hops increases, the same signal is sent over a higher number of hops and thus it will be more likely that the Willies can detect the communication. 

The performance curves of the schemes as a function  of path-loss exponent intersect at some points. That is, for a small number of system nodes, the covert throughput when  $\alpha$ is small  is higher, but as the number of system nodes increases, the covert throughput when $\alpha$ is larger becomes higher. The reason is that when the number of system nodes is small, Alice and Bob have very few choices of nodes to construct a path, and thus the optimal path might not be able to  avoid Willies effectively. Hence, the throughput when $\alpha$ is small is higher because smaller $\alpha$ leads to smaller signal attenuation and thus higher throughput. 
But when the number of system nodes becomes larger, Alice and Bob have many choices to construct a path that can avoid Willies. And, when the path-loss exponent is large, the higher attenuation of the environment makes the transmission of each relay local (because each relay has a smaller broadcast range), which  helps the system nodes to avoid Willies more effectively. Thus, when the number of system nodes is higher, we have a higher covert throughput for larger path-loss exponents.

\begin{figure}
	\centering
	\includegraphics[width=.5\textwidth]{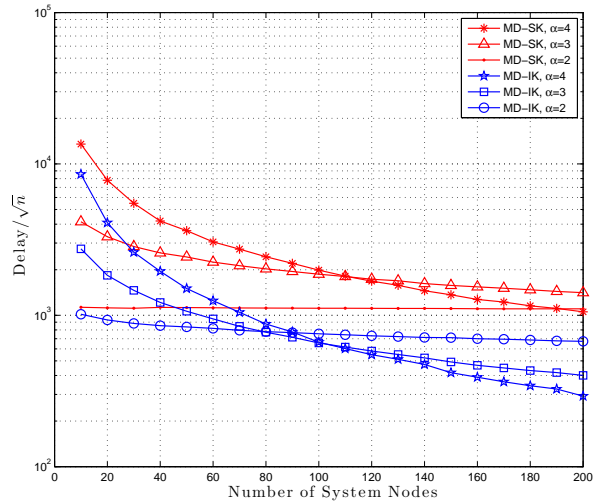}
	\caption{End-to-end delay versus number of system nodes, when  30 Willies are present in the network and the covertness factor $\delta=0.05$.}
	\label{fig:delay_vs_num_system}
\end{figure}

Fig. \ref{fig:delay_vs_num_system}
shows the end-to-end delay of the MT-SK and MT-IK algorithms  versus the number of system nodes, when $\delta=0.05$ and 30 Willies are present in the network. It can be seen that the performance of MT-IK is better than the performance of MT-SK.
As the number of system nodes increases, the end-to-end delay decreases because the  routing algorithm has a larger set from which to choose the relays so as to minimize the end-to-end delay.

For all algorithms, when the number of system nodes is small, for smaller $\alpha$ we have better performances. 
The reason is that a smaller path-loss exponent means less attenuation and thus higher throughput. However, as the number of system nodes increases, the optimal path can take advantage of more system nodes to take detours and avoid Willies. In this case, the performance of the  algorithms when the path-loss exponent is large is better compared to when the path-loss exponent is small. The reason is that when the path-loss exponent is large the effect of each Willie is local and taking detours can improve the performance to a greater extent.

\begin{figure}
	\centering
	\includegraphics[width=.5\textwidth]{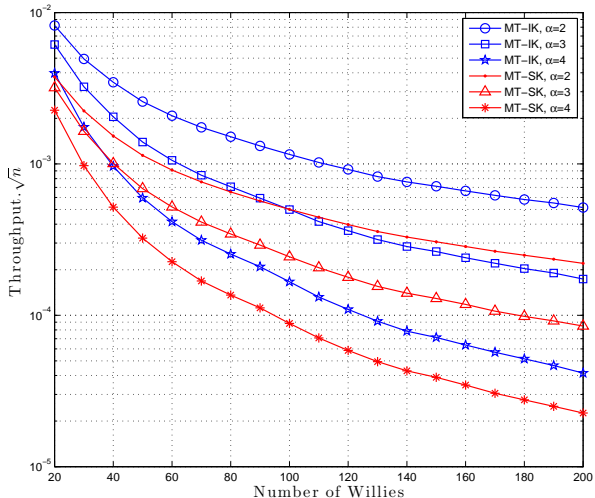}
	\caption{Throughput versus the number of Willies, when  30 system nodes are present in the network and the covertness factor $\delta=0.05$.}
	\label{fig:rate_vs_num_willies}
\end{figure}

\textbf{Effect of Number of Willies.}
The effect of the number of Willies on the maximum covert throughput achieved by MT-SK and MT-IK is shown in Fig. \ref{fig:rate_vs_num_willies}. In this figure, $\delta=0.05$ and 30 system nodes are present in the network.
With both algorithms, as the number of Willies increases, the maximum throughput of covert communication decreases. In all situations considered in this figure, the performance of the MT-IK algorithm is better than the performance of MT-SK algorithm, as expected.

\begin{figure}
	\centering
	\includegraphics[width=.5\textwidth]{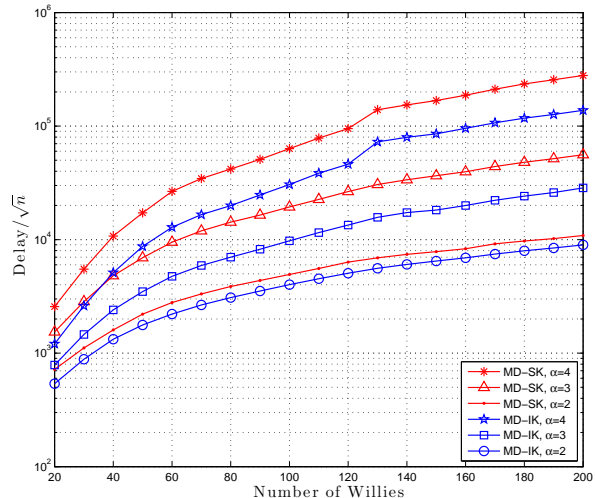}
	\caption{End-to-end delay versus the  number of Willies, when  30 system nodes are present in the network and the covertness factor $\delta=0.05$.}
	\label{fig:delay_vs_num_willies}
\end{figure}

The end-to-end delay versus the number of Willies is shown in Fig. \ref{fig:delay_vs_num_willies}, when 30 system nodes are present in the network and $\delta=0.05$. As expected, the end-to-end delay of transmission from source to destination increases as the number of Willies increases, because with more Willies the throughput of communication at each link becomes smaller,  and the optimum path should take more detours to avoid Willies,  resulting in a larger number of hops. It can be seen that MD-IK always has a smaller end-to-end delay than MD-SK.

\section{Concluding Remarks}\label{sec:conc}

In this paper,  multi-hop covert communication over an arbitrary network  in an AWGN environment and in the presence of multiple collaborating Willies has been considered.
 We developed maximum throughput and minimum end-to-end delay routing algorithms for \textit{a single key  for all relays} approach, and  for \textit{independent keys at the relays} approach.  
 We have shown that using these multi-hop algorithms improves the performance of traditional one-hop covert communication from Alice to Bob substantially. 
 Each proposed routing algorithm is straightforward to implement, and finds the optimal path in polynomial time in the size of the network.
 
 We have shown mathematically and via  simulations that for different network parameters the  performance (throughput and delay) of routing algorithms with independent keys is better compared to that of routing algorithms with a single key for all relays.
Note that the better performance of routing with independent keys is gained at the expense of a higher number of key bits used for covert communication.
As mentioned in Section \ref{sec:key}, the long key sequences can be generated from short keys pre-shared between the system nodes.
In this paper, we pictured a scenario where the  system nodes are co-located and share  keys, and then they are distributed over the network area and use their pre-shared keys for covert communication.  
An exciting direction for future work is to consider  covert wireless key distribution.

 \appendices
 \section{}
 \label{ap:0}
 	The relative entropy between $\mb{Q}_1$ and $\mb{Q}_0$ can be written as,
 	\begin{align}
 	\mb{D}(\mb{Q}_1\|\mb{Q}_0)&=\mb{E}_{\mb{Q}_1}\{\log\mb{Q}_1-\log\mb{Q}_0\}.\label{eq:d1}
 	\end{align}	
 %	where $h(\mb{Q}_1)$ is the differential entropy of $\mb{Q}_1$. Since the zero-mean Gaussian distribution
 	%maximizes differential entropy among all distributions of the same variance, we have,
 	Then,
 	\begin{align*}
 	&\mb{E}_{\mb{Q}_1}\{\log\mb{Q}_1\}\\
 	&=\frac{\dim(\Sigma_1)}{2}\log(2\pi)+\frac{1}{2}\log|\Sigma_1|\\
 	&\quad+\frac{1}{2}\mb{E}_{\mb{Q}_1}\{(x-\mu_1)\Sigma_1^{-1}(x-\mu_1)\}\\
 	&=\frac{\dim(\Sigma_1)}{2}\log(2\pi)+\frac{1}{2}\log|\Sigma_1|+\frac{1}{2}\Tr{\Sigma_1^{-1}\Sigma_1}\\
 	&\quad+\frac{1}{2}(\mu_1-\mu_1)\Sigma_1^{-1}(\mu_1-\mu_1)\numberthis\label{eq:h1}\\
 	&=\frac{\dim(\Sigma_1)}{2}\log(2\pi)+\frac{1}{2}\log|\Sigma_1|+\frac{1}{2}\dim (\Sigma_1),\numberthis\label{eq:h2}
 	\end{align*}
 	where (\ref{eq:h1}) follows from \cite[Section 8.2.2]{petersen2008matrix},
 	and,
 	\begin{align*}
 	&\mb{E}_{\mb{Q}_1}\{\log\mb{Q}_0\}\\
 	&=\mb{E}_{\mb{Q}_1}\Big\{-\frac{\dim(\Sigma_0)}{2}\log(2\pi)-\frac{1}{2}\log|\Sigma_0|\\
 	&\quad-\frac{1}{2}(x-\mu_0)\Sigma_0^{-1}(x-\mu_0)\Big\}\\
 	&=-\frac{\dim(\Sigma_0)}{2}\log(2\pi)-\frac{1}{2}\log|\Sigma_0|\\
 	&\quad-\frac{1}{2}\mb{E}_{\mb{Q}_1}\{(x-\mu_0)\Sigma_0^{-1}(x-\mu_0)\}\\
 	&=-\frac{\dim(\Sigma_0)}{2}\log(2\pi)-\frac{1}{2}\log|\Sigma_0|-\frac{1}{2} \Tr\{\Sigma_0^{-1}\Sigma_1\}\\
 	&\quad+\frac{1}{2}(\mu_1-\mu_0)\Sigma_0^{-1}(\mu_1-\mu_0), \numberthis\label{eq:e1}
 	\end{align*}
 	where  (\ref{eq:e1}) follows from \cite[Section 8.2.2]{petersen2008matrix}.
 	Combining (\ref{eq:d1}), (\ref{eq:h2}) and (\ref{eq:e1}),
 	\begin{align*}
 	\mb{D}(\mb{Q}_1\|\mb{Q}_0)&= \frac{1}{2}\Bigg( \Tr\{\Sigma_0^{-1}\Sigma_1\}+(\mu_1-\mu_0)\Sigma_0^{-1}(\mu_1-\mu_0)\\
 	&\quad+\log\frac{|\Sigma_0|}{|\Sigma_1|}-\dim(\Sigma_1)\Bigg).
 	\end{align*}

\section{}
\label{ap:1}
Here we calculate the relative entropy between $\mb{Q}_1$ and $\mb{Q}_0$ of Section \ref{sec:covertness_AF}. The relative entropy of two multivariate Gaussian random variable $\mb{Q}_1=\mc{N}(\mu_1,\Sigma_1)$ and $\mb{Q}_0=\mc{N}(\mu_0,\Sigma_0)$ is,
\begin{align}\label{eq:ap_1}
\mathbb{D}(\mathbb{Q}_1\|\mathbb{Q}_0)&
\nonumber= \frac{1}{2}\Big( \Tr(\Sigma_0^{-1}\Sigma_1)+(\mu_0-\mu_1)^T\Sigma_0^{-1}(\mu_0-\mu_1)\\
&-\dim(\Sigma_0)-\ln\left( \frac{|\Sigma_1|}{|\Sigma_0|}\right)  \Big).
\end{align}
 We use the same approach as in \cite[Appendix]{soltani2014covert} to calculate each term of (\ref{eq:ap_1}).
 The first term can be written as,
\begin{align*}
\Tr(\Sigma_0^{-1}\Sigma_1)&=n\Tr(S^{-1}(S+UU^T))\\
&=n\Tr(I_{HM\times HM}+S^{-1}UU^T)\\
&=nHM+n\sum_{i=1}^H\sum_{k=1}^M\frac{P_i}{\sigma_{W_k}^2 d_{i,k}^{\alpha}}.\numberthis
\end{align*}
The second term vanishes because $\mu_0=\mu_1=0$. The third term is,
\[\dim(\Sigma_0)=\dim(S\otimes I_{n\times n})=nHM.\]
The forth term can be calculated as,
\begin{align*}
|\Sigma_0|&=|S\otimes I_{n\times n}|\\
&=|S|^n| I_{n\times n}|^{HM}\\
&=|S|^n.\numberthis
\end{align*}
and,
\vspace{-10pt}
\begin{align*}
|\Sigma_1|&=|(S+UU^T)\otimes I_{n\times n}|\\
&=|S+UU^T|^n| I_{n\times n}|^{HM}\\
&=|S|^n|I+S^{-1}UU^T|^n\\
&=|S|^n(I+U^TS^{-1}U)^n\\
&=|\Sigma_0|\left(1+ \sum_{i=1}^H\sum_{k=1}^M\frac{P_i}{\sigma_{W_k}^2d_{i,k}^{\alpha}}\right) ^n.\numberthis
\end{align*}
\vspace{-10pt}
Thus,
\begin{align*}
&\mathbb{D}(\mathbb{Q}_1\|\mathbb{Q}_0)\\&
=\frac{1}{2}\Big( \Tr(\Sigma_0^{-1}\Sigma_1)+(\mu_0-\mu_1)^T\Sigma_0^{-1}(\mu_0-\mu_1)\\
&\quad-\ln \frac{|\Sigma_1|}{|\Sigma_0|}-\dim(\Sigma_0) \Big) \\
&=\frac{n}{2}\Big( HM+\sum_{i=1}^H\sum_{k=1}^M\frac{P_i}{\sigma_{W_k}^2 d_{i,k}^{\alpha}}\\
&\quad-\ln\Big( 1+ \sum_{i=1}^H\sum_{k=1}^M\frac{P_i}{\sigma_{W_k}^2d_{i,k}^{\alpha}}\Big) -HM\Big) \\
&=\frac{n}{2}\Big( \sum_{i=1}^H\sum_{k=1}^M\frac{P_i}{\sigma_{W_k}^2 d_{i,k}^{\alpha}}	\\
&\quad-\ln\Big( 1+ \sum_{i=1}^H\sum_{k=1}^M\frac{P_i}{\sigma_{W_k}^2d_{i,k}^{\alpha}}\Big) \Big) 
\end{align*}

	\section{}
	\label{ap:2}
	In this appendix we present the solution of the following optimization problem:
		\begin{align}\label{eq:ap_maxmin3}
	\nonumber	\max\left( \min_i C_i\right) ,\;i=1,\ldots,H\\
		 \text{s.t.}\;\sum_{\ell_i\in\Pi}\left(\sum_{W_k\in\mc{W}} \frac{P_i}{\sigma^2_{W_k}d_{i,k}^{\alpha}}\right)^2\leq\gamma_2.
		\end{align}
	We claim that $\max\left( \min_i C_i\right)$ in (\ref{eq:ap_maxmin3}) is obtained when all  links $\ell_i\in\Pi$  have the same throughput. 
	Let us define,
	\begin{align}
	g_i=\left( \sum_{W_k\in\mc{W}}\frac{2\sigma_{i}^2d_{i}^{\alpha}}{\sigma^2_{W_k}d_{i,k}^{\alpha}}\right) ^2.
	\end{align}
	Hence, we should maximize $\min_i C_i$ such that $\sum_i C_i^2 g_i\leq\gamma_2$.
	Suppose  $C_{(1)}=\min_i C_i,\; i=1,\ldots,H$.
	We have,
	\begin{align*}
	\gamma_2\geq\sum_iC_i^2g_i\geq C_{(1)}^2\sum_ig_i,
	\end{align*} 
	and thus,
	\begin{align*}
	C_{(1)}\leq\sqrt{\frac{\gamma_2}{\sum_ig_i}}.
	\end{align*}
	Setting
	\begin{eqnarray}
	C_1=\cdots=C_{H}=\sqrt{\frac{\gamma_2}{\sum_i g_i}},
	\end{eqnarray}
	proves the claim.

	\bibliographystyle{ieeetran}
	\bibliography{references}

\end{document}